\documentclass[11pt]{article}
\usepackage{moriond,epsfig}

\def\be{\begin{equation}}
\def\ee{\end{equation}}
\def\bea{\begin{eqnarray}}
\def\eea{\end{eqnarray}}

\begin{document}

\newcommand{\half}{{1\over2}}
\newcommand{\nad}{n_{\rm ad}}
\newcommand{\niso}{n_{\rm iso}}
\newcommand{\fiso}{f_{\rm iso}}
\newcommand{\ii}{\'{\'i}}
\newcommand{\bk}{{\bf k}}
\newcommand{\Ocdm}{\Omega_{\rm cdm}}
\newcommand{\ocdm}{\omega_{\rm cdm}}
\newcommand{\OM}{\Omega_{\rm M}}
\newcommand{\OB}{\Omega_{\rm B}}
\newcommand{\oB}{\omega_{\rm B}}
\newcommand{\OL}{\Omega_\Lambda}
\newcommand{\cltt}{C_l^{\rm TT}}
\newcommand{\clte}{C_l^{\rm TE}}
\newcommand{\calR}{{\cal R}}
\newcommand{\calS}{{\cal S}}
\newcommand{\Rrad}{{\cal R}_{\rm rad}}
\newcommand{\Srad}{{\cal S}_{\rm rad}}

\vspace*{4cm}
\title{NEW CONSTRAINTS ON DARK ENERGY}

\author{Alessandro Melchiorri}

\address{Dipartimento di Fisica, Universita' di Roma ``La Sapienza'',
INFN Sezione di Roma, Ple Aldo Moro 5, 00185, Roma, Italy.}

\maketitle\abstracts{New Cosmic Microwave 
Background, Galaxy Clustering and Supernovae type Ia data 
are increasingly  constraining the dark energy component
of our Universe.
While the cosmological constant scenario remains consistent with these 
new tight constraints, the data does not rule out the possibility 
that the equation of state parameter is less than $-1$}

\section{Introduction}

The recent results of precision cosmology have been extremely important since 
they provide an excellent agreement with our theoretical picture of 
the cosmos, incorporating the standard model of structure formation, 
the inflationary prediction of flatness, the presence of cold dark matter 
and an amount of baryonic matter consistent with Big Bang Nucleosynthesis 
constraints (see e.g.~\cite{spergel},  ~\cite{riess},  ~\cite{tegmark2}).
The price-tag of this success story concerns a very puzzling consequence: 
the evolution of the universe is dominated by a mysterious form of energy, 
$X$, coined dark energy, (an unclustered negative pressure component of the 
mass-energy density), with a present-day energy density fraction 
$\Omega_X \simeq 2/3$ and equation of state parameter
(pressure over energy density ratio) $w_X \equiv p_X/\rho_X \sim -1$
(or even  $w_X < -1$, see ~\cite{mmot}). 

\noindent This discovery may turn out to be 
one of the most important contribution to physics in our generation.
Hence it is especially important to consider all possible scheme
for dark energy.

\noindent A true cosmological constant $\Lambda$ may be at works here.
However, as it is well known, is difficult to associate the 
small observed value of the cosmological constant $\rho_\Lambda
\sim (10^{-3}eV)^4$ with vacuum fluctuations in scalar field theories,
which, for example, for bosonic and fermionic fields would
led to an effective cosmological constant of 
$\rho_\Lambda\sim10^{76}GeV^4$, i.e. of $123$ orders of magnitude larger.
Moreover, the cosmological constant immediately introduces 
a ``why now'' problem, since an extreme fine-tuning of initial
conditions is required in order to obtain $\rho_\Lambda\sim\rho_m$ 
{\it today}: already at redshift $z \sim 2$ the cosmological constant is 
subdominant, while at the time of the electroweak phase transition
  $\rho_\Lambda /\rho_m \sim 10^{-55}$.

\noindent Systematics in the data are most probably under control:
combined analyses of CMB, LSS and SN-Ia data yield 
$\Omega_\Lambda=0.74\pm0.04$, i.e.a more than $14 \sigma$'s 
detection.
The SN-Ia alone is highly inconsistent with $\Omega_\Lambda=0$ if
one consider flat universes or open with $\Omega_M > 0.1$.
The CMB data alone is also inconsistent with $\Omega_\Lambda=0$
unless one considers closed models with $\Omega_M\sim 1.3$ and
a very low Hubble parameter $h \sim 0.4$ which, again, are 
incompatible with several complementary datasets.

\noindent Assuming modifications to the model of structure formation
which are not connected with a new form of energy, like, 
for example, a contribution from isocurvature perturbations,
doesn't seems able to mimic $\Lambda$ or a dark energy
component (see e.g. ~\cite{trotta}).

\section{Alternatives to $\Lambda$}

A complete treatment of the possible contenders to the dark energy throne
can be found in several and excellent recent reviews 
(see e.g. ~\cite{peebles},~\cite{sahni},
~\cite{padmanabhan},~\cite{trodden2}).
The important point is that dark energy candidates have an
equation of state parameter which can be different from $-1$ 
and varies with time compared to that of a cosmological constant
which remains fixed at $w_{\Lambda}=-1$. Thus, observationally 
distinguishing a time variation in
the equation of state or finding $w_X$ different from $-1$ will
rule out a pure cosmological constant as an explanation for the data,
but be consistent with a dynamical solution.

\noindent Here let me mention few models, 
according to the expected values of their equation of state:

\subsection{Topological Defects, $-1/3 \ge w_X \ge -2/3$}

Dark energy can receive contributions from topological defects
produced at phase transitions in the early universe (see e.g.
\cite{vilenkin},\cite{dkm}). However, despite a well established
theoretical framework, topological defects have not been thoroughly 
explored due to technical difficulties in the numerical simulations.
More recently, a plausible version of dark energy
made of a frustrated network of domain walls
was proposed by \cite{bucher} (see also \cite{eichler}).
These models have several appealing features:
Firstly, topological defects are ubiquitous in field theory
and unavoidable in models with spontaneously broken symmetries.
Second, the scale of spontaneous symmetry breaking responsible for
the walls is expected to lie in the $10-100$ $KeV$ range
and can arise naturally in supersymmetric theories (\cite{friedland}).
Finally a firm phenomenological prediction can be
made for domain walls models: an equation of state strictly
$-1/3 \ge w_X \ge -2/3$ (see e.g. \cite{friedland}).
These models are therefore predictive in the value of the equation
of state parameter and distinguishable from
a cosmological constant even at zero order on $w_X$, (while, for example, 
scalar field models can also produce $w_X\sim-1$ although they differ from 
a cosmological constant which in the first order variation has 
$\dot w_X = 0$).

\subsection{Scalar Fields - Quintessence $-2/3 \ge w_X \ge -1$}

It is entirely possible that a dynamic mechanism is giving rise to the 
observed acceleration of the present Universe.
Some of the popular proposed candidates to explain the observations are a 
slowly-rolling scalar field, ``quintessence''
~\cite{Wetterich:fm}-\cite{Caldwell:1997ii}, or a ``k-essence'' 
scalar field with non-canonical kinetic terms in the Lagrangian 
~\cite{Armendariz-Picon:1999rj}-\cite{Chiba:1999ka}.
An important property of these models is that, since the equation
of state is time dependent, the fine tuning (``why now'') problem
can be in principle alleviated.
Several models have been proposed and a complete study of all the
related potentials goes well beyond the present $6$-pages work.
As mentioned, the most general prediction is a value for the equation 
of state $w(z)$ that differs from unity and varies with redshift $z$. 
A second way to distinguish between scalar field candidates is to measure 
the sound speed of the dark energy component that affects the perturbations 
in its energy distribution. The sound speed in many models of quintessence 
is equal to the speed of light, however can be different from $c$, for
example, in k-essence models, where it varies,
triggered by the transformations in the background
equation of state.

\subsection{Phantom or Super-Quintessence,  $-1 \ge w_X$}

As we will see in the section, the present data does not rule
out but even slightly suggest $w_X < -1$. 
Scalar field models with such equation of state (known as 
``phantom'' or super-quintessence models) deserve a separate discussion
since they cannot be achieved by scalar fields with positive 
kinetic energy term. The limitation to $w_X>-1$ is indeed 
a theoretical consideration motivated, for example, by
imposing on matter (for positive energy densities) the null energy condition,
which states that $T_{\mu\nu}N^{\mu}N^{\nu}>0$ for all null 4-vectors 
$N^{\mu}$. Such energy conditions are often
demanded in order to ensure stability of the theory.
However, theoretical attempts to obtain $w_X <-1$ have been
considered~\cite{Caldwell:1999ew,Schulz:2001yx,parker,frampton,Ahmed:2002mj}.
Unstable at quantum level, a careful analysis of their potential instabilities 
has been performed in~\cite{trodden}.
Moreover, the expansion factor of a universe dominated by phantom
energy diverges in a finite amount of cosmic time, culminating in
a future curvature singularity (Big rip \cite{bigrip} or
Big smash \cite{McInnes} phase).

\subsection{Chaplyngin gases  $w_X=-1$ today, 
$w_X=0$ yesterday}

The Chaplyngin Gas (CG) (see e.g. \cite{charlie}) provides an interesting
 possibility for an unified picture of dark energy and dark matter
since such component interpolates in time between dust ($w_X=0$)and a 
cosmological constant ($w_X=-1$), with an intermediate behavior as 
$p=\alpha \rho$. Perturbations of this fluid are stable on small scales, 
but behave in a very different way with respect to standard quintessence.
Analysis of the effect of those perturbations on CMB and LSS data, in 
particular, have strongly constrained CG, disfavouring it as an unified 
dark matter candidate (see e.g. \cite{bean}).

\section{Analysis of the current data.}

In order to bound $w_X$, we consider a template of flat, adiabatic,
$X$-CDM models computed with CMBFAST~\cite{sz}. We sample the
relevant parameters as follows:
$\Omega_{cdm}h^2 = 0.05,...0.20$, in steps of  $0.01$;
$\Omega_{b}h^2 = 0.015, ...,0.030$ (motivated by Big Bang 
Nucleosynthesis), in steps of  $0.001$, $\Omega_{Q}=0.0, ..., 0.95$,
in steps of  $0.05$ and $w_X=-3.0,...,-0.4$ in steps of $0.04$,
assumed as constant with redshift. 
For most of the dynamical models on the market, the assumption of a 
piecewise-constant equation of state is a good approximation for an 
unbiased determination of the effective equation of state (\cite{wang0})
\begin{equation}
w_{\rm eff} \sim \frac{\int w_X(a) \Omega_X(a) da}{\int \Omega_X(a) da}
\end{equation}
predicted by the model.
Hence, if the present data is not compatible with a constant
$w_X=-1$, it may be possible to discriminate between a cosmological
constant and a dynamical dark energy model.

\noindent The value of the Hubble constant in our database is not an
independent parameter, since it is determined through the flatness 
condition.We adopt the conservative top-hat bound $0.45 < h < 0.85$
and we also consider the $1\sigma$ constraint on the Hubble
parameter, $h=0.71\pm0.07$, obtained from Hubble Space Telescope
(HST) measurements~\cite{freedman}. 

\noindent We allow for a reionization of the intergalactic medium by
varying the Compton optical depth parameter
$\tau_c$ over the range $\tau_c=0.05,...,0.30$ in steps of $0.02$.
 
\noindent For the CMB data we use the recent temperature and 
cross polarization results from the WMAP satellite
(\cite{Bennett:2003bz}) using the method explained in 
(\cite{Verde:2003ey}) and the publicly available code
on the LAMBDA web site.
As in \cite{mmot}, we further include the results from the 
BOOMERanG-98~\cite{ruhl}, DASI~\cite{halverson}, MAXIMA-1~\cite{lee},
CBI~\cite{pearson}, VSAE~\cite{grainge} experiments by using the
 publicly available correlation matrices and 
window functions.
We consider $7 \%$, $10 \%$, $4 \%$, $5 \%$, $3.5 \%$  and $5 \%$
Gaussian distributed calibration errors for the BOOMERanG-98, DASI, 
MAXIMA-1, VSA, and CBI experiments respectively.

\noindent In addition to the CMB data we also consider
the real-space power spectrum
of galaxies in the 2dF $100$k and SLOAN first year 
galaxy redshift survey using the
data and window functions of the analysis of ~\cite{thx} and ~\cite{tegmark2}.
We restrict the analysis to a range of scales over which the fluctuations 
are assumed to be in the linear regime ($k < 0.1 h^{-1}\rm Mpc$).
When combining with the CMB data, we marginalize over a bias $b$
for each dataset considered to be an additional free parameter.

\noindent We finally incorporate constraints obtained
from the luminosity measurements of Type Ia supernovae (SN-Ia)
from ~\cite{riess} using the GOLD dataset and
again evaluating the likelihoods assuming a constant equation of state.

\begin{figure}[t]
\begin{center}
\includegraphics[scale=0.4]{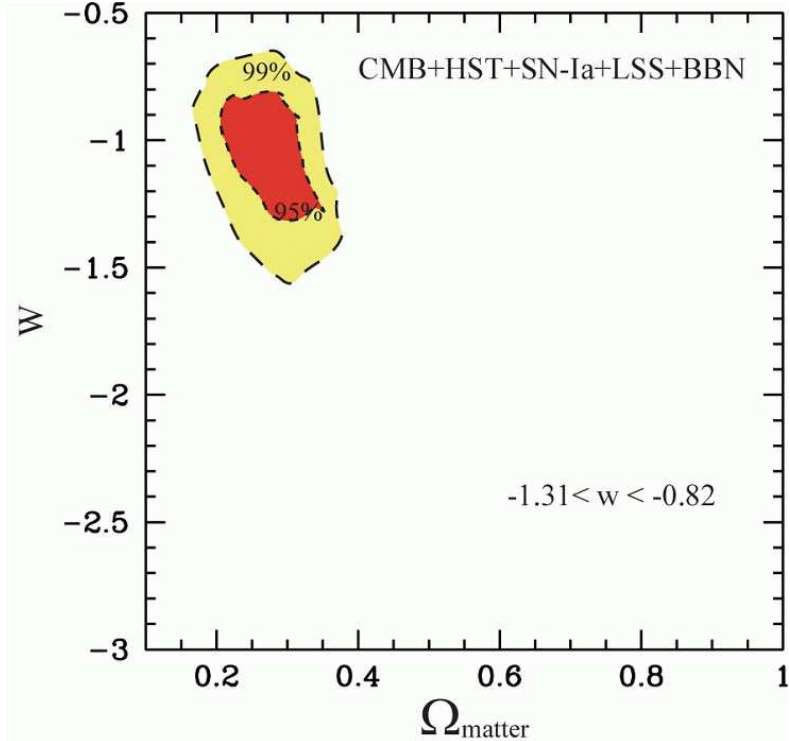}
\end{center}
\caption{Likelihood contours in the ($\Omega_M$, $w_X$) plane for the
joint CMB+HST+SN-Ia+LSS analysis described in the text. We take the best-fit
values for the remaining parameters.
The contours correspond to 0.05 and 0.01 of the peak value of the
likelihood, which are the 95\% and 99\% confidence levels respectively.}
\label{figo1}
\end{figure}

\noindent In Figure~\ref{figo1} we plot the likelihood contours in the
($\Omega_M$, $w_X$) plane from our joint analyses of CMB+SN-Ia+HST+LSS
data. As we can see, there is strong supporting evidence for dark energy.
A cosmological constant with $w_X=-1$ is in good agreement with all the
data. However the $2$-$\sigma$ confidence levels
are  $-1.32 < w_X <-0.82$ with a best-fit value of $w_X \sim 1.04$,
slightly preferring ``phantom'' models.

\noindent While the analysis rules out topological defects as dark energy,
it is important to note that this result is almost completely due to
the inclusion of the Supernovae Type-Ia dataset.
Topological defects can provide a good fit to the WMAP data for a different 
choice of priors with ``lower'' values of the Hubble parameter ($h<0.65$), 
(as indicated by Sunyaev-Zeldovich and time delays for gravitational lensing 
observations), and ``higher'' values of the matter density 
($\Omega_m > 0.35$), (in agreement with recent measurements of the 
temperature-luminosity relation of distant clusters observed with the 
XMM-Newton satellite) (see ~\cite{conversi}).

\noindent A cosmological constant is compatible with our analysis 
but this result may be biased by the assumption of a constant with redshift 
equation of state. However, analysis of recent supernovae data,
while still compatible with an evolution of $w_X$ 
(see the contribution of M. Giavalisco), are 
not providing an evidence for such variation.

\section{Conclusions}

We have demonstrated that, even by applying the most current constraints
on the dark energy equation of state parameter $w_X$, there is much
uncertainty in its value. Interestingly, there is a distinct possibility
that it may lie in the theoretically under-explored region $w_X <-1$. 
An observation of a component to the cosmic energy budget with $w_X <-1$
would naturally have significant implications for fundamental physics.
Further, depending on the asymptotic evolution of $w_X$, the fate of
the observable universe~\cite{Starkman:1999pg}-\cite{Huterer:2002wf} may be
dramatically altered, perhaps resulting in an instability of the
spacetime~\cite{trodden} or a future singularity.
 
\noindent If we are to understand definitively whether dark energy 
is dynamical, and if so, whether it is consistent with $w_X$ less than or 
greater than $-1$, we will need to bring the full array of cosmological 
techniques to bear on the problem. An important contribution to this effort 
will be provided by direct searches for supernovae at both intermediate 
and high redshifts~\cite{SNAP}.
Other, ground-based observations~\cite{LSST} will allow complementary analyses,
including weak gravitational lensing~\cite{Huterer:2001yu} and large scale
structure surveys~\cite{Hu:1998tk} to be performed.
 
\noindent At present, however, while the data remain
consistent with a pure cosmological constant $\Lambda$.

\section*{Acknowledgments}

I would like to thank the organizers for this great conference 
on the italian Alps. I am indebted with my collaborators: 
Luca Conversi, Laura Mersini, Carolina Odman, Mark Trodden and Joe Silk.

\end{document}